\documentclass[aps,prl,twocolumn,groupedaddress,amsmath,letter,showpacs]{revtex4}

\usepackage{graphicx}
\usepackage{array}
\usepackage{dcolumn}
\usepackage[french]{babel}
\usepackage[latin1]{inputenc}
\begin{document}

%\preprint{}

\title{The origin of the $E_+$ transition in GaAsN alloys}


\author{V. Timoshevskii, N. Madini, M. C\^ot\'e, and R. Leonelli}
%\thanks{}
\affiliation{D\'{e}partement de Physique et Regroupement Qu\'{e}b\'{e}cois
sur les Mat\'{e}riaux de Pointe (RQMP) \\
Universit\'{e} de Montr\'eal, Case Postale 6128, Succursale Centre-ville,
Montr\'eal, Qu\'ebec, H3C 3J7 Canada}

\date{\today}

\begin{abstract}
Optical properties of GaAsN system with nitrogen concentrations in the range of 0.9-3.7\% are studied by full-potential LAPW method in a supercell approach. The $E_+$ transition is identified by calculating the imaginary part of the dielectric function. The evolution of the energy of this transition with nitrogen concentration is studied and the origin of this transition is identified by analyzing the contributions to the dielectric function from different band combinations. The $L_{1c}$-derived states are shown to play an important role in the formation of the $E_+$ transition, which was also suggested by recent experiments. At the same time the nitrogen-induced modification of the first conduction band of the host compound are also found to contribute significantly to the $E_+$ transition. Further, the study of several model supercells demonstrated the significant influence of the nitrogen potential on the optical properties of the GaAsN system.
\end{abstract}

%\keywords{}

\maketitle


%% ========================================
%% Short Introduction, ...and brief results
%% ========================================
GaAs$_{1-x}$N$_x$ alloys have attracted much attention of both experimentalists and theoreticians due to their unusual physical properties. Among those features, observed in the typical nitrogen concentration regime of 0.5\%-3.0\%, one should mention the anomalous band gap reduction, which makes this system technologically important for such electronic devices as infrared diode lasers \cite{Kondow96} and multijunction solar cells \cite{Geisz98}. Another striking feature of this system is the appearance of a new composition-dependent optical transition, called $E_+$, detected by electro-reflectance measurements for the samples with nitrogen concentrations of $\geq$0.8\% \cite{Perkins99, Francoeur03}. This peak is located at about 0.4-0.8 eV above the conduction band (CB) minimum (denoted as $E_-$) and shifts upward in energy with increase of nitrogen concentration \cite{Perkins99, Perkins01, Francoeur03}.

Several physical mechanisms have been proposed to describe these nitrogen-induced changes: two-level band anticrossing (BAC) model \cite{Shan99,Skierb01}, disorder-allowed $\Gamma$-X-L coupling \cite{Jones99, Mattila99}, and the formation of the impurity band \cite{Zhang00}. In the frames of the BAC model the $E_+$ state naturally appears as the high-energy solution of a two-state Hamiltonian, describing the interaction of the localized nitrogen state with the extended CB states of the host matrix compound \cite{Oreilley02}. However, the first-principles \cite{Jones99,Gonzalez01,Wang01,Gorczyca02} as well as empirical pseudopotential \cite{Mattila99,Kent01prl,Kent01prb} calculations failed to validate the BAC model. On the other hand, based on these calculations, the intraband coupling model has been proposed. In this model the $E_+$ peak is caused by a disorder-allowed transition from the valence band (VB) maximum to the $L$-point of the conduction band. In particular, it was proposed that the energy position of the $E_+$ transition is a configuration-weighted average of a nitrogen impurity state and CB $L$-state, denoted as $L_{1c}$ \cite{Mattila99}. However, to our knowledge, there are no studies of the GaAsN system, which identify and clarify the origin of the $E_+$ transition based on direct first-principles calculations of the optical properties of this system.

In this Letter we report the results of theoretical study of the optical properties of GaAsN system using state-of-the-art \textit{ab-initio} method based on density functional theory (DFT). The $E_+$ transition is identified by direct calculation of the dielectric function of the system for nitrogen concentrations of 0.9-3.7\%. Its origin is further clarified by studying band-to-band contributions to the dielectric function. To understand the effects of atomic ordering, the supercells of different symmetry are constructed and studied separately. Our results show that the $E_+$ transition is formed by the interplay of two contributions: the $\Gamma$-point transitions to the $L_{1c}$-derived states of the GaAs matrix, and the $L$-point transitions to the impurity-modified first conduction band of the system. Further, we study the changes in these contributions, caused by different impurity distribution, and show the importance of nitrogen potential in the formation of the $E_+$ transition.

% =======================
% Calculation techniques
% =======================
Our calculations were performed within the local density approximation (LDA) and generalized gradient approximation (GGA) to the density functional theory \cite{HK-KS}. Four supercells were constructed to describe GaAs$_{1-x}$N$_x$ system with nitrogen concentrations in the range of $x=$0.0093-0.037. In 64-atom and 216-atom supercells nitrogen atoms form a simple cubic (SC) lattice, while in 54-atom and 128-atom supercells a face-centered cubic (FCC) impurity sublattice is formed. A pseudopotential planewave method as implemented in ABINIT \cite{abinit} code was employed for full geometry optimization of the supercells. This included structural relaxation both with respect to the volume of the unit cell and to the internal coordinates of the atoms. A 16 Ry basis cutoff and 2$\times$2$\times$2 Monkhorst and Pack \cite{MP} sampling of the Brillouin zone (BZ) showed good results for structural relaxation. The LDA exchange-correlation functional of Perdew and Zunger \cite{PZ81} has been used for calculations.

Having obtained geometrical characteristics of the supercells, we proceeded further with a highly-accurate linearized augmented plane wave method (LAPW) as implemented in WIEN2K program package \cite{Wien2k}. This method belongs to the family of all-electron DFT-type methods and does not use any pseudopotentials. Band structures and optical properties of the model systems have been calculated using the GGA exchange-correlation functional of Engel and Vosko \cite{EV93}. This functional significantly improves the values of the band gap and effective electron and hole masses for GaAs \cite{Persson01} as well as for GaAsN \cite{Bentoumi04} systems. For calculation of optical properties a dense 20$\times$20$\times$20 \textbf{k}-grid has been used \cite{grids}. Having obtained selfconsistently the crystal wavefunctions $\vert\textbf{k}n\rangle$ and band energies $E_{\textbf{k}n}$ for $n$ bands of the system, the imaginary part of the dielectric function $\varepsilon_2(\omega)$ can be calculated by summing direct transitions over the BZ, weighted with corresponding matrix element of the transition probability:
\begin{eqnarray*}
\varepsilon_2^{ii}(\omega) & = & \displaystyle  \frac{Ve^2}{2\pi\hbar m^2 \omega^2}\int d^3k \sum_{nn'}\vert\langle\textbf{k}n\vert p_{i}\vert\textbf{k}n' \rangle\vert^2 \\
& & \times f_{\textbf{k}n}(1-f_{\textbf{k}n'})\delta(E_{\textbf{k}n'}-E_{\textbf{k}n}-\hbar\omega).
\end{eqnarray*}
Here \textbf{p}=$-i\hbar\nabla$ is the momentum operator, and $f_{\textbf{k}n}$ is the Fermi distribution function, which ensures that only transitions from occupied to unoccupied states are included in summation.

% =======================
% Discussion of results
% =======================
\begin{figure}[t]
  \includegraphics*[width=0.9 \linewidth]{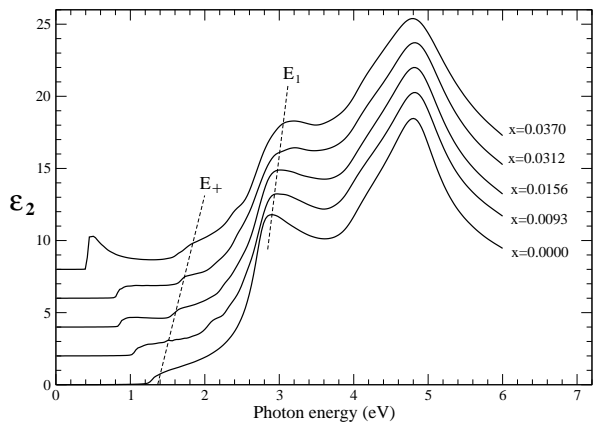} \\
  \caption{\label{epsilon} Calculated imaginary part of the dielectric function for GaAs$_{1-x}$N$_x$ systems, normalized per unit volume. A progressive Lorentzian smearing with full width at half maximum (FWHM) of 0.01($\hbar\omega$)$^2$ has been applied.}
\end{figure}
We present in Fig.\ref{epsilon} calculated imaginary part of the dielectric function ($\varepsilon_2$) for GaAsN systems with nitrogen concentrations of 0.93, 1.56, 3.13, and 3.70\%. A calculated $\varepsilon_2$ for pure GaAs is also presented for sake of comparison. All main features, observed experimentally for GaAs and GaAsN systems, are well-reproduced by theoretical calculations. The common features of the dielectric function for the pure and doped systems include the presence of the fundamental absorption edge $E_0$, as well as two main maxima, namely peaks $E_1$ and $E_2$. The origin of these features is well-understood and described in the literature \cite{Yu01}. The fundamental absorption edge is due to transitions between valence band maximum and conduction band minimum, which occur in the center of the Brillouin zone. The $E_1$ peak is related to direct transitions along the eight equivalent $\Gamma-L$ directions of the BZ. And, finally, the absolute maximum $E_2$ contains contributions from the direct transitions occurring near $X$-point of the BZ.

It is clearly seen that with increase of nitrogen doping the fundamental edge moves down to lower energies. This reflects the effect of the nitrogen-caused band gap reduction, and agrees well with experimental observations \cite{Skierb01, Bentoumi04}. We also notice that the $E_2$ peak hardly moves with increase of the nitrogen contents, while the $E_1$ transition moves slightly to higher energies. The main feature, which makes the spectra of GaAs and GaAsN systems substantially different, is the appearance of a new nitrogen-induced peak, which we attribute to the $E_+$ transition, observed experimentally. With increase of nitrogen content the $E_+$ peak shifts to higher energies faster than the $E_1$ transition.

\begin{figure}[t]
  \includegraphics*[width=0.9 \linewidth]{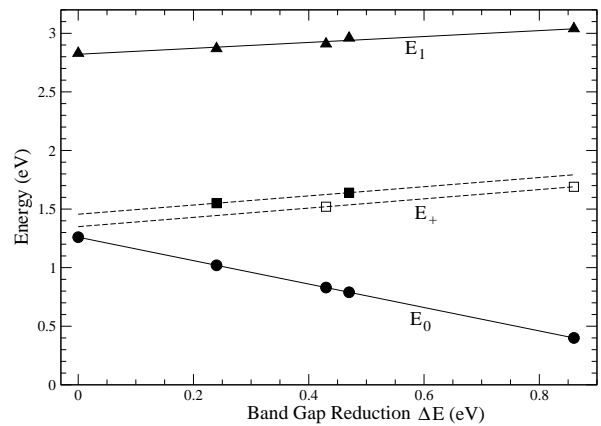} \\
  \caption{\label{energies} Optical transition energies vs band gap reduction: the $E_0$, $E_+$, and $E_1$ transitions are presented.} 
\end{figure}

To represent our results quantitatively, we plot in Fig.\ref{energies} the optical transition energies as a function of band gap reduction $E_0$(GaAs)-$E_0$(GaAsN) with the energies measured relative to the VB maximum. In this representation the $E_0$ transition is a straight line with a slope of -1.0. The energy evolution of the $E_1$ peak can also be well-approximated by a straight line with a slope of +0.25. This slope is close to +0.22, experimentally obtained by Shan \textit{etal} \cite{Shan00} for four different nitrogen concentrations with further parametrization of the results within the BAC model. 

While for $E_0$, $E_1$ and $E_2$ the exact transition energies can be easily determined directly from the dielectric function, this task becomes much more difficult for the $E_+$ peak. This transition is clearly visible only for 64- and 128- atom supercells. For the system with the highest nitrogen concentration of 3.7\% this transition is smeared to large extend probably due to strong N-N interactions, while for the smallest concentration of 0.93\% it is hardly visible due to a reduced oscillator strength. By comparing the dielectric function for 128- and 64- atom supercells with corresponding energy spectra at the $\Gamma$-point, we find that the $E_+$ transition energy for these structures exactly corresponds to the position of the $a_1(L_{1c})$ singlet. At this point we are making an important assumption, which will be later verified by a detailed analysis. We assume that for the other two structures the $E_1$ peak also originates from a $\Gamma$ transition to the $a_1(L_{1c})$ singlet. This assumption is also supported by the fact that for our model system, containing 0.93\% of nitrogen, this singlet is located at 0.53 eV above the conduction band minimum, which is in good agreement with experimental value of 0.57 eV, obtained for the $E_+$ peak for the samples with 1\% of nitrogen \cite{Perkins01}.

Now we plot in Fig.\ref{energies} the energies of the $E_+$ transitions within the same representation. Linear approximations, performed separately for SC- and FCC- types of supercells, showed practically the same slope of +0.40, which is smaller than experimental value of +0.67, obtained by Perkins \textit{et al} \cite{Perkins99} for the samples with four different nitrogen concentrations. Extrapolation to the dilute limit gives the average value of $E_+-E_0(x=0)=145$ meV, which is close to the experimental interval of 150-180 meV, obtained for the energy position of nitrogen level in GaAs \cite{Kent01prb}. However, such a detailed comparison should be done with a caution due to the fact that only two points are available for each type of structures. 

\begin{figure}[t]
 \includegraphics*[width=1.0 \linewidth]{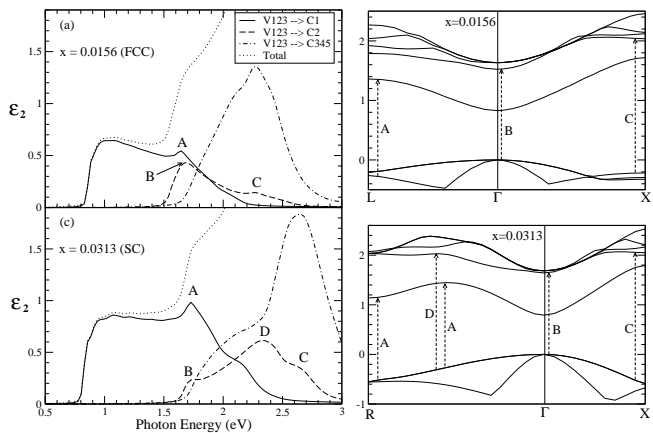}
  \caption{\label{eps_decomp} (a) and (c): Contributions to the imaginary part of the dielectric function from different band combinations, calculated for 128- and 64- atom supercells. (b) and (d): Calculated electronic bandstructure of these supercells showing the various direct transitions, responsible for the features in $\varepsilon_2$.}
\end{figure}

We now try to understand the origin of the $E_+$ transition in the GaAs$_{1-x}$N$_x$ system. The expression, used for the calculation of $\varepsilon_2(\omega)$, gives us a natural possibility to calculate the contributions to the dielectric function from diffrent band combinations. Figure \ref{eps_decomp}(a,c) represents this decomposition of $\varepsilon_2$, performed for the systems with 1.56\% and 3.13\% of nitrogen, taking into consideration only three top valence and five lowest conduction bands. Only these combinations were found to be responsible for the formation of the dielectric function in the energy interval 0.0-2.5 eV, which includes the $E_+$ transition.

Let us first concentrate on the system with 1.56\% of nitrogen (Fig. \ref{eps_decomp}a). We consider three contributions to the dielectric function: transitions to the first conduction band (C$_1$), to the second conduction band (C$_2$), and cumulative transitions to the conduction bands 3, 4 and 5 (C$_{345}$). We observe that the transitions to C$_{345}$ do not contribute to the near-edge features of $\varepsilon_2$. The results show that the $E_+$ feature is formed by contributions of two transitions, marked A and B, which both are energetically located near 1.65 eV, but have completely different origin. Analyzing the energy positions of peaks A and B, and keeping in mind that they come from the transitions to different conduction bands C$_1$ and C$_2$, we are able to uniquely connect these transitions with corresponding saddle points in the bandstructure of the system. Figure \ref{eps_decomp}b represents the bandstructure of the 128-atom supercell, showing three top valence and five lowest conduction bands. The results show that the peak A is formed by the transitions from the the valence bands to the first conduction band near the $L$-point of the BZ, while the peak B is due to the transitions to the second conduction band near the $\Gamma$-point. The fact that these two transitions are very close in energy and have significant intensities, leads to the formation of a well-defined feature in the spectrum, which we attribute to be the $E_+$ transition. The transitions near the $X$-point of the BZ contribute to the spectrum in the form of the peak C. It is formed by the transitions to the second conduction band, but it is twice smaller than peak B in intensity, and is located at higher energies.

It was found that large configuration-induced bandgap fluctuations are present in the GaAsN systems both in cases of one-dimensional nitrogen chains \cite{Yacoub00}, as well as at relatively low nitrogen concentrations \cite{Bentoumi04}. It seems interesting to study the influence of impurity ordering on the formation of the $E_+$ transition. So, we proceed further with the same decomposition analysis of $\varepsilon_2$ for the 64-atom supercell, where the nitrogen atoms form a simple cubic sublattice. The contributions of the C$_1$, C$_2$ and C$_{345}$ bands are represented in Fig. \ref{eps_decomp}c, and the band structure with corresponding optical transitions is shown in Fig. \ref{eps_decomp}d. The results show than the $E_+$ transition is also formed by a combination of peaks A and B, but the relative intensities of these peaks change. Comparing Fig. \ref{eps_decomp}a and Fig. \ref{eps_decomp}c we observe that the contribution of peak B decreases, while the one of peak A increases. The increase of the contribution of the band C$_1$ for the SC-type supercell can be explained by the appearance of the additional saddle point in the $\Gamma-R$ valley. The saddle point of the same type appeares also in the C$_2$ band, which leads to the formation of additional peak D, which is not observed in the FCC-type structure.

We now conclude our study by exploring the role of impurity potential in the process of the formation of the $E_+$ transition. Besides splitting the $L_{1c}$-states into a singlet and a triplet, the introduction of nitrogen also leads to a strong perturbation of the first conduction band of the host compound \cite{Kent01prb}. These two effects could be caused by both the geometrical deformations near the impurity atom, and by the nitrogen potential itself. In this last part of our study we are making an attempt to verify the influence of these two effects on the formation of the $E_+$ transition.

To gain further insight we performed calculations of the electronic structure for two model systems with subsequent analysis of different contributions to the dielectric function.  We have first taken the 128-atom supercell with the fully relaxed geometry and replaced nitrogen atom back to arsenic. Thus, in this model we have removed the effects of the impurity potential, staying only with the effects of structural relaxation. The results show two dramatical changes in the electronic structure and optical properties of the system: the band gap of this model structure opens to the value of 1.29 eV, which is close to 1.26 eV, obtained for the GaAs structure, and the $E_+$ transition is no longer present in the spectrum. The contribution analysis demonstrated the reduction of the amplitudes of peaks A and B to practically negligible values. Moreover, in spite of local geometrical deformations, the splitting of $L_{1c}$ states is now only 0.06 eV, which is almost twice lower than 0.11 eV, observed for the N-doped supercell. This demonstrates that local mechanical deformations introduce only small perturbations to the GaAs system, which are not responsible neither for dramatic band gap reduction, nor for the appearance of the $E_+$ transition.

These conclusions are supported by our results, obtained for the second model structure. In this case we have taken the N-doped 128-atom supercell, but stayed within the geometry of the GaAs structure. This model allows us to observe only the effects of the nitrogen potential without any influence of impurity-caused deformations. The decomposition analysis showed that both peaks A and B, which are responsible for the formation of the $E_+$ transition, are present in the spectrum. The differences from the relaxed structure include the larger splitting of the $L_{1c}$ states (0.17 eV) and shifts in the energy positions of peaks A and B. We also observe opening of the band gap up to 1.13 eV. However, this value is smaller than the one, obtained for the previous model system, where only the absense of the nitrogen potential caused a dramatic opening of the band gap to the one of the GaAs structure.

In conclusion, we have presented theoretical study of the optical properties of the GaAsN system. The origin of the experimentally observed $E_+$ transition is identified by the analysis of the dielectric function, calculated \textit{ab-initio} for four different supercells. Our results show that the main contribution to the $E_+$ peak is given by  $\Gamma$-point transitions from the VB maximum to the $a_1(L_{1c})$-singlet, which originates from a nitrogen-induced splitting of the CB $L$-point. At the same time, nitrogen-induced modifications of the first CB are found to give rise to additional $L$-point direct transitions. Being close in energy to $\Gamma$-point transitions, they are shown to contribute significantly to the formation of the $E_+$ peak. Finaly, local geometrical deformations are found to be only a second order effect in the process of the formation of the $E_+$ transition, and the influence of the impurity potential itself is shown to be responsible for the $E_+$ transition and dramatic band gap reduction in GaAsN.

\bibliography{GaAsN_Eplus}

\end{document}